\documentclass[pra, reprint, twocolumn, amsmath,amsfonts,amssymb]{revtex4-1}
\usepackage{graphicx}
\usepackage{amsmath,amsfonts,amssymb}
\usepackage[colorlinks=true,linkcolor=blue]{hyperref}
\usepackage{color}
\usepackage{bm}
\usepackage{threeparttable}
\usepackage[mathscr]{eucal}
\usepackage{picture}

\usepackage{tikz}
\usetikzlibrary{quantikz}

\usepackage{color}
\usepackage{bm}
\definecolor{grey}{rgb}{0.7,0.7,0.7}

\usepackage{fancyvrb}
\usepackage{ulem}

\usepackage[T1]{fontenc}
\usepackage{braket}
\allowdisplaybreaks
\begin{document}
\title{Improved algorithms of quantum imaginary time evolution for ground and excited states of molecular systems}
\author{Takashi Tsuchimochi}
\email{tsuchimochi@gmail.com}
\affiliation{Graduate School of System Informatics, Kobe University, 1-1 Rokkodai-cho, Nada-ku, Kobe, Hyogo 657-8501 Japan}
\affiliation{Japan Science and Technology Agency (JST), Precursory Research for Embryonic Science and Technology (PRESTO), 4-1-8 Honcho Kawaguchi, Saitama 332-0012 Japan}
\author{Yoohee Ryo}
\affiliation{Graduate School of Science, Technology, and Innovation, Kobe University, 1-1 Rokkodai-cho, Nada-ku, Kobe, Hyogo 657-8501 Japan}
\author{Seiichiro L. Ten-no}
\affiliation{Graduate School of System Informatics, Kobe University, 1-1 Rokkodai-cho, Nada-ku, Kobe, Hyogo 657-8501 Japan}

\begin{abstract}
Quantum imaginary time evolution (QITE) is a recently proposed quantum-classical hybrid algorithm that is guaranteed to reach the lowest state of system. In this study, we present several improvements on QITE, mainly focusing on molecular applications. We analyze the derivation of the underlying QITE equation order-by-order, and suggest a modification that is theoretically well founded. Our results clearly indicate the soundness of the here-derived equation, enabling a better approximation of the imaginary time propagation by a unitary. We also discuss how to accurately estimate the norm of an imaginary-time-evolved state, and applied it to excited state calculations using the quantum Lanczos algorithm. Finally, we propose the folded-spectrum QITE scheme as a straightforward extension of QITE for general excited state simulations. The effectiveness of all these developments is illustrated by noiseless simulations, offering the further insights into quantum algorithms for imaginary time evolution.
\end{abstract}
\maketitle

\section*{Introduction}
Future applications of quantum computers are diverse because of its expected capability of solving complex problems that are difficult with classical computers. While quantum computing is expected to become a game changer for everyday technology such as machine-learning\cite{Dunjko18, Zhang20} and cryptography\cite{Pirandola20}, another significant potential application is quantum chemistry simulation for material design owning to the inherent nature of quantum entanglement in electronic structures\cite{Cao19, McArdle20}.

For the noisy intermediate-scale quantum computer, various quantum-classical hybrid algorithms have been developed to determine the ground states and also excited states of chemical Hamiltonians. Many of them are based on the variational quantum eigensolver (VQE)\cite{Peruzzo14}, which, using classical computers, optimizes parameters in a fixed quantum circuit. VQE has been extensively studied\cite{Wecker15,Kandala17,Wang19} for molecules\cite{OMalley16, McClean16, Shen17, Grimsley19} and recently extended to solid states\cite{Cerasoli20,Fan21,Yoshioka22}.

However, the classical optimization of VQE in a high-dimensional, non-linear parameter space poses a challenge to determine the ground state without being trapped in a local minimum. Recently, several algorithms based on imaginary time evolution (ITE) have emerged to circumvent the gradient-based parameter optimization\cite{Motta20, McArdle19, Yeter-Aydeniz20, Gomes20, Gomes21, Sun21, Huang22, Amaro22}. ITE is able to transform an arbitrary state to the (nearly) exact ground state, and has been historically applied to fermion systems on the basis of Monte Carlo simulations\cite{Anderson75, Anderson76, Blankenbecler81, Linden90, Varney09, Zhang03, Ohtsuka08, Booth09}. Quantum ITE (QITE), developed by Motta and co-workers\cite{Motta20}, performs an approximate unitary evolution that mimics the imaginary time propagation. The convergence of QITE is accelerated by quantum Lanczos diagonalization (QLanczos) that expands a Krylov subspace with each time step. The potential of these approaches has been demonstrated for spin models and small molecular systems such as H$_2$, by both simulation and real quantum devices\cite{Motta20, Yeter-Aydeniz20, Yeter-Aydeniz21}. Several authors have further extended QITE by employing an adaptive approach\cite{Gomes21} and merging imaginary time steps to a single unitary\cite{Yeter-Aydeniz20, Gomes20}. However, most previous studies focused on relatively simple model systems and could not guarantee the accuracy and applicability of QITE for more general problems. 

To expand QITE's applicability to chemical systems, in this work, we will first introduce a sparse representation of the QITE approximation based on fermion operators. Using this scheme, we will focus on both ground and excited states of molecules, and assess the scalability of the method to more complex chemical systems.

In QITE, the unitary evolution that approximates the imaginary time evolution is determined by solving a set of linear equations to minimize the distance between the two (virtually) evolved states. In this study, we revisit the derivation and propose a modified equation based on an order-by-order analysis. As shown below, our equation clearly provides a considerably higher fidelity of the unitary evolved state especially for large imaginary time step $\Delta\beta$. It, consequently, enables faster convergence of the QITE simulation, and thus saves the required quantum resource. This improvement is considered important for strongly correlated systems, as they require a relatively longer imaginary time in general. 

In addition, we discuss the potential of QLanczos in determining excited states. Note that QLanczos was tested for excited states by Yeter-Aydeniz and co-workers, showing promising results for very simple systems such as H$_2$\cite{Yeter-Aydeniz20, Yeter-Aydeniz21}. We were also able to reproduce these results in our own simulation. However, for more complex (larger) systems, QLanczos gives rather unsatisfactory results because of the difficulty in  estimating the required matrix elements accurately. Hence, we introduce refinements aiming for improved descriptions of excited states. Furthermore, the folded-spectrum propagator is proposed to directly tackle general eigenstates, which are difficult to treat with QLanczos. 

This article is organized as follows. In the Results section, we introduce a chemistry-inspired framework for QITE and derive its linear equation based on our own perspective. We then perform ground state calculations to analyze the accuracy of different algorithms and methods. We also propose improved evaluations of QLanczos matrix elements and the folded-spectrum propagator, which are both applied to excited state calculations. Finally, we conclude this work in the Discussion section.

\section*{Results}\label{sec:Theory}
\subsection*{Quantum imaginary time evolution for chemical systems}\label{sec:QITE}
Before presenting our results, we first recap the QITE algorithm. The imaginary time evolution of an arbitrary state $|\Phi_0\rangle$ that has a non-zero overlap with the ground state $|\psi_0\rangle$ is expressed as,
\begin{align}
	|\psi_0\rangle 
	&\propto  \lim_{n\rightarrow \infty}\left(e^{-\Delta\beta \hat H }\right)^n |\Phi_0\rangle
\end{align}
where $\Delta\beta$ is the imaginary time step and $\hat H$ the target Hamiltonian. Note that we have introduced the Trotter approximation such that $\Delta\beta$ is sufficiently short. The propagator is not unitary and therefore the essential idea of QITE is that the action of $e^{-\Delta\beta \hat H}$ to an intermediate, normalized quantum state $|\Phi^{(\ell)}\rangle = e^{-\ell\Delta\beta \hat H} |\Phi_0\rangle / \sqrt{\langle \Phi_0|e^{-2\ell\Delta\beta\hat H}|\Phi_0\rangle}$ with the integer $\ell$ is approximated by some unitary $e^{-i\Delta\beta\hat A}$, which can be efficiently implemented on a quantum circuit. Hereafter, we will drop the superscript $(\ell)$ when its presence is obvious, for simplicity.

Although such a unitary $e^{-i\Delta\beta\hat A}$ definitely exists, it is unclear how one can efficiently build it. Originally, QITE assumed a local structure in each term of the Hamiltonian, which was exploited to construct $\hat A$ as a linear combination of Pauli strings,
\begin{align}
	\hat A = \sum_{\mu} a_\mu \hat \sigma_\mu.\label{eq:Am}
\end{align}
Here, $\hat \sigma_\mu \in \{I, X, Y, Z\}^{\otimes D}$ where the correlation domain $D$ determines the representability of $e^{-i\Delta\beta\hat A}$ to approximate the imaginary time propagation with each local term in the Hamiltonian. The real coefficients $a_\mu$ are determined by minimizing the norm error between the two states obtained from ITE and the unitary, which can be processed with a classical computer (see the Supplementary Information). This procedure is repeated for all the local Hamiltonian terms for each imaginary time step $\Delta\beta$. To accelerate the convergence, the quantum Lanczos (QLanczos) method was proposed, in which an effective Hamiltonian defined within the Krylov subspace $\{|e^{-\ell\Delta\beta\hat H}|\Phi\rangle\}$ is diagonalized. These methods have been tested for 1D Heisenberg model and other local models with successful results\cite{Motta20, Sun21}.

For molecular systems, we consider the aforementioned local treatment is not particularly suitable, because the definition of $D$ can become ambiguous. We note that a chemical Hamiltonian is also considered local and sparse because the interaction is limited to between two bodies:
\begin{align}
	\hat H = \sum_{pq} h_{pq} c^\dag_p c_q + \frac{1}{2}\sum_{pqrs} (pr|qs) c^\dag_p c^\dag_q c_s c_r \label{eq:H}
\end{align} 
where $p,q,r,s$ indicate spin-orbitals, and $c^\dag_p$ and $c_p$ are creation and annihilation operators. 
Here, $h_{pq}$ and $(pq|rs)$ are the standard notation for conventional molecular integrals\cite{Szabo}.
We can therefore introduce Fermionic-QITE that uses anti-Hermitian fermionic operators to construct the chemistry-inspired $\hat A$,
\begin{subequations}
	\begin{align}
	\hat A &= \hat A_1 + \hat A_2 +\cdots \label{eq:A=A1+A2} \\
	\hat A_1 &= \sum_{pq} t_{pq} (c^\dag_p c_q - c^\dag_q c_p) \\
	\hat A_2 & = \frac{1}{4}\sum_{pqrs} t_{pqrs} (c^\dag_p c^\dag_q c_s c_r - c^\dag_r c^\dag_s c_q c_p)
\end{align}\label{eq:A}\end{subequations}
where we have assumed the particle-number symmetry in the Hamiltonian and ${\bf t}$ is purely imaginary. $\hat A_k$ comprises  the $k$-rank excitation and de-excitation operators, and Eq.~(\ref{eq:A}) is appropriately transformed to the qubit basis as Eq.~(\ref{eq:Am}). Then, it is easily seen that the expansion is complete in the sense that $e^{-i \Delta \beta \hat A}$ can generate any number-preserving state in principle. Furthermore, since the short time evolution $e^{-\Delta \beta \hat H} = 1 - \Delta \beta \hat H + O(\Delta \beta^2)$ is viewed as one and two particle substitutions of the reference state $|\Phi\rangle$ to first order, one can truncate $\hat A$ after the second term of Eq.~(\ref{eq:A=A1+A2}) to obtain a good approximation. The unitary with such $\hat A = \hat A_1 +\hat A_2$ is known as unitary coupled-cluster with generalized singles and doubles (UCCGSD), which has been shown to be satisfactorily accurate in the context of VQE\cite{Lee19}, and we call QITE using this {\it ansatz} ``UCCGSD-based QITE'' in this article. In passing, Gomes et al. explored the UCCSD {\it ansatz} for QITE\cite{Gomes20}, which only includes the excitations with respect to the Hartree-Fock (HF) vacuum\cite{Barkoutsos18, Romero19}.
 

\subsection*{Corrected equation for QITE}
Having discussed our scheme for chemical Hamiltonians above, the goal of QITE is to find the unitary $e^{-i\Delta\beta\hat A}$ that approximates the imaginary time propagation on a given state $|\Phi\rangle$. This can be achieved by minimizing the following function $F({\bf a})$. In contrast to the original work, we perform the order-by-order analysis, and establish the equation to the second-order of $\Delta\beta$, 
\begin{align}
F({\bf a}) & =\|\frac{1}{\sqrt{c}}e^{-\Delta\beta \hat H}|\Phi\rangle - e^{-i\Delta\beta\hat A}|\Phi\rangle\|^2  \nonumber\\
	&= 2 - \frac{ 2}{\sqrt{c}}{\rm Re}\langle \Phi|e^{-\Delta\beta \hat H}  e^{-i\Delta\beta\hat A}|\Phi\rangle\nonumber\\
& = const.  +  \Delta\beta^2 \Bigl( \langle \Phi | \hat A^2 | \Phi\rangle - i\langle \Phi|\left[\hat H, \hat A\right]| \Phi\rangle \Bigr)  \nonumber\\
&+ O(\Delta\beta^3) \label{eq:F}
\end{align}
Here, $c = \langle \Phi| e^{-2\Delta\beta\hat H}|\Phi\rangle$ is the squared norm of the imaginary time-evolved state.
It should be noted that the problem is similar to the maximization of the fidelity between the above-mentioned two states. A quite similar formula to the second equality was proposed by Benedetti et al. in the context of the variational scheme for time evolution\cite{Benedetti21}. Also note that we have expanded $\frac{1}{\sqrt{c}}$ in terms of $\Delta \beta$ as well, but keeping this term constant would not change our result because the first-order term of the overlap ${\rm Re}\langle \Phi|e^{-\Delta\beta \hat H}  e^{-i\Delta\beta\hat A}|\Phi\rangle$ is a constant (i.e., independent of {\bf a}), see the Supplementary Information. 
Hence, we minimize the simpler function
\begin{align}
	f({\bf a}) = \sum_{\mu\nu} \langle \Phi| \hat \sigma_\mu \hat \sigma_\nu |\Phi\rangle a_\mu a_\nu - i\sum_{\mu}  \langle \Phi|\left[\hat H, \hat \sigma_\mu\right]|\Phi\rangle a_\mu,\label{eq:f}
\end{align}
resulting in the following equation:
\begin{align}
	{\bf M}{\bf a} + {\bf b} = {\bf 0}  \label{eq:Aa+b}
\end{align}
where
\begin{align}
	M_{\mu\nu} &=2 {\rm Re} \langle \Phi|\hat\sigma_\mu \hat\sigma_\nu|\Phi\rangle \label{eq:Amat}\\
	b_\mu &= {\rm Im} \langle \Phi | \left[\hat H, \hat\sigma_\mu\right] |\Phi\rangle \label{eq:b}
\end{align}
Eq.~(\ref{eq:b}) differs from the original derivation\cite{Motta20},
\begin{align}
	b_\mu^{\rm prev} = \frac{2}{\sqrt{c}} {\rm Im}\langle \Phi|\hat H \hat\sigma_\mu |\Phi\rangle \label{eq:borg}
\end{align}
in that the factor $\frac{1}{\sqrt{c}}$ is {\it not} present. Furthermore, we suggest to exploit the commutator form in Eq.~(\ref{eq:b}) with $\left[\hat H, \sigma_\mu\right]$, which helps reduce the complexity arising in $\hat H\sigma_\mu$. In passing, it is remarkable to regard $b_\mu$ as the energy derivative around the reference state,
\begin{align}
	b_\mu = \frac{\partial}{\partial (\Delta\beta a_\mu)} \left.\langle\Phi| e^{i\Delta \beta \hat A} \hat H e^{-i\Delta \beta \hat A} |\Phi\rangle \right|_{{\bf a} = {\bf 0}},
\end{align}
which clarifies the physical meaning of Eq.~(\ref{eq:Aa+b}): QITE can be viewed as a part of the natural gradient descent algorithm, noting that $\langle \Phi|\sigma_\mu|\Phi\rangle = 0$ and thus {\bf M} is equivalent to the Fubini-Study metric tensor\cite{Stokes20, Gacon21}. The algorithm is considered to have converged when the gradient {\bf b} becomes zero.

The metric {\bf M} has a null space because of the redundancy in $\{\sigma_\mu|\Phi\rangle\}$, and therefore it has infinite solutions for {\bf a}. In other words, $\|{\bf a}\|_2$ can become quite large, which triggers numerical instabilities (we are assuming the Trotter approximation in $e^{-i\Delta\beta \hat A}$); in this case, regularization is performed.

Let us now consider and demonstrate the consequence of the factor of $\frac{1}{\sqrt{c}}$ in $b_\mu$ (but not in $M_{\mu\nu}$), as derived in the original proposal of Ref.~[\onlinecite{Motta20}] and used in other studies\cite{Gomes20, Yeter-Aydeniz20, Sun21, Yeter-Aydeniz21, Huang22}. 
 According to our analysis in the Supplementary Information, its expansion order is inconsistent in that $c$ contains the $O(\Delta\beta)$ dependence. Therefore, in the original equation, to our understanding, the update is incorrectly scaled by $\frac{1}{\sqrt{c}}$ ($< 1$ in many cases), leading to a significant deceleration of the convergence if $\Delta\beta$ is large. For small molecules like H$_2$, this behavior is virtually invisible and thus may have been overlooked. However, it is anticipated to have a greater influence on large molecules because $c$ increases exponentially with the energy.

\begin{figure}
	\includegraphics[width=80mm]{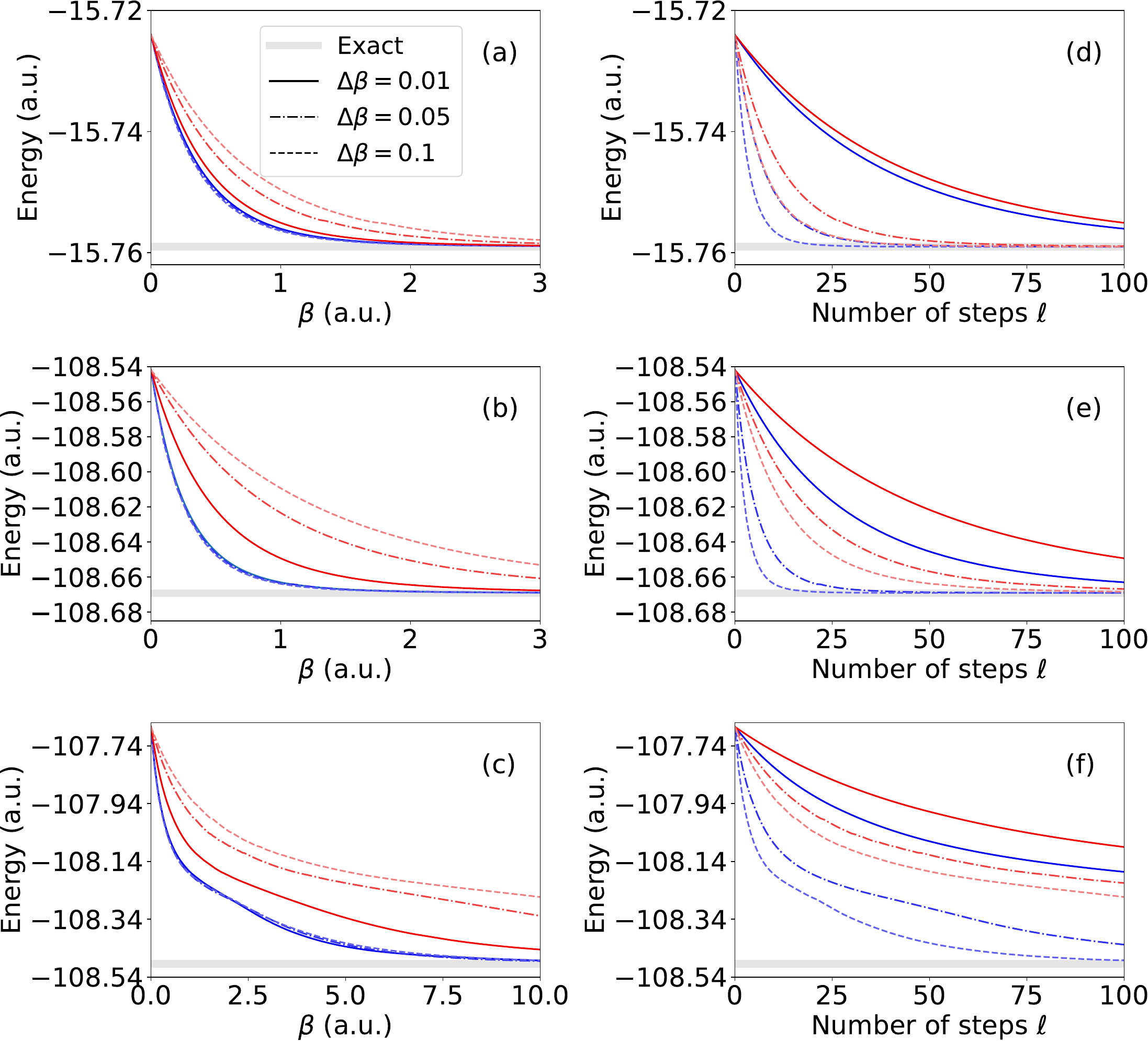}
	\caption{Energy convergence of the UCCGSD-based QITE for different imaginary time step sizes $\Delta\beta$. Red and blue curves indicate the previous and present equations, respectively. Thick gray lines are the corresponding exact energy. Convergence as a function of $\beta$ for BeH$_2$ (a) and N$_2$ with the equilibrium bond length (b) and the stretched bond length (c). Convergence as a function of the imaginary time steps $\ell$ of BeH$_2$ (d) and N$_2$ with the equilibrium bond length (e) and the stretched bond length (f).} \label{fig:gs}
\end{figure}

The test systems used here are the linear BeH$_2$ molecule at equilibrium (R$_{\rm e} = 1.334$ {\AA}) and the N$_2$ molecule at equilibrium and dissociation (R$_{\rm e} = 1.098$ and R$_{\rm dis} = 2.5$ \AA, respectively). The initial state is set to HF, and the UCCGSD-based QITE is used to propagate the initial state with different time steps, $\Delta\beta =  0.01, 0.05$, and $0.1$ {\it a.u.}. We have used the STO-6G Gaussian basis set to represent atomic orbitals. The Be 1$s$ orbital and the N 1$s$ and 2$s$ orbitals are frozen. For the dissociating N$_2$, we have used the L2 regularization to stabilize the linear equation with a regularization constant of $10^{-7}$.

In Figs.~\ref{fig:gs}(a), (b), and (c), we plotted the energy convergence against the imaginary time $\beta$ for BeH$_2$ and N$_2$ at R$_{\rm e}$ and R$_{\rm dis}$, respectively. It is noteworthy that, in all these calculations, the energy error from the exact value at a large $\beta$ limit is less than 1 mHartree. However, it is evident that the convergence profiles for QITE with the three different time steps significantly  differ from each other when simulated based on the original derivation Eq.~(\ref{eq:borg}), see the red curves. The smaller $\Delta\beta$, the faster convergence is achieved with respect to the total imaginary time $\beta$. However, in total, the smaller $\Delta\beta$ results in more steps compared to the larger one. This result is illustrated well in Figs.~\ref{fig:gs}(d), (e), and (f), which show the energy convergence with respect to the actual number of times the linear equation is solved. Importantly, the deteriorated convergence behavior with the original derivation is more pronounced for N$_2$ than for BeH$_2$, because the energy is one order of magnitude larger, and so is $c$.  Also importantly, the slow convergence is far evident for N$_2$ at the dissociation limit as shown in Figs.~\ref{fig:gs}(c) and (f) (compared with (b) and (e)), indicating the potential difficulty in applying QITE to strongly correlated systems.

\begin{figure}
	\includegraphics[width=85mm]{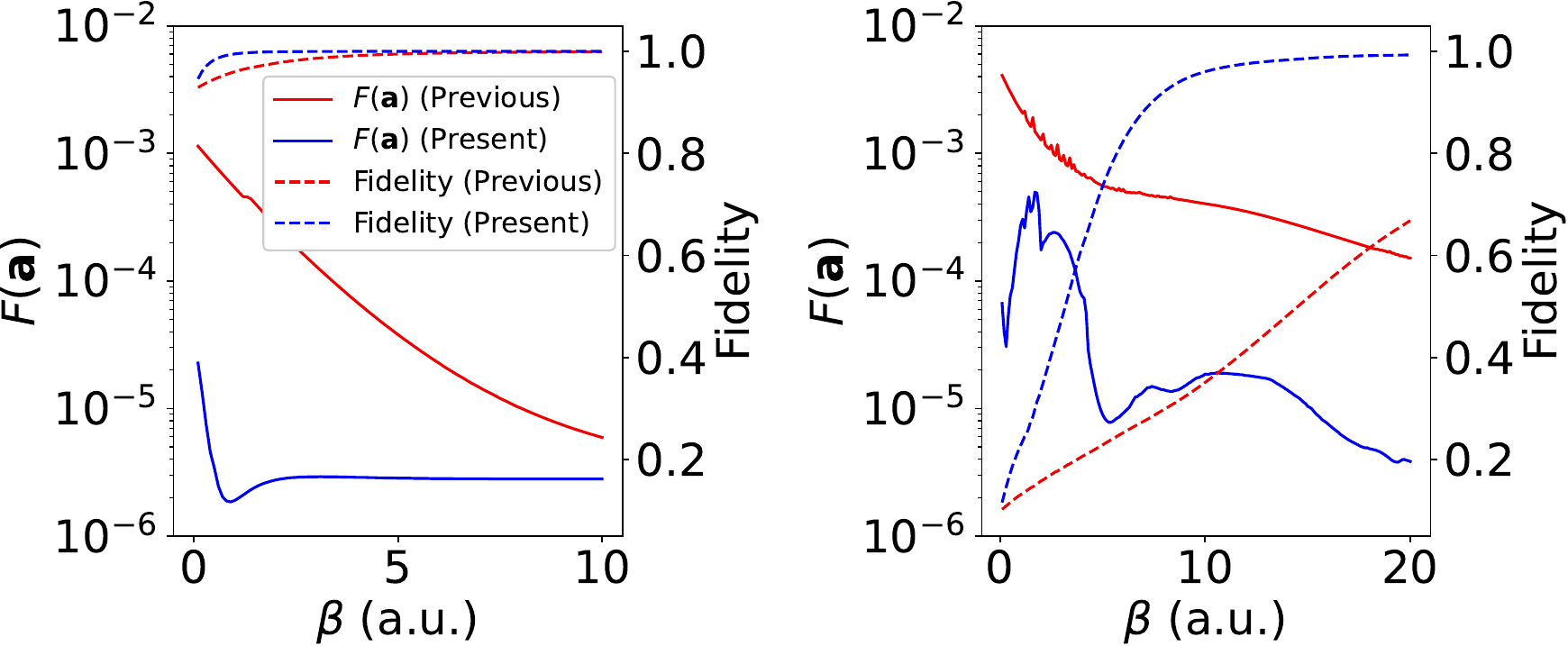}
	\caption{Difference indicator $F$ between $\frac{1}{\sqrt{c}} e^{-\Delta\beta \hat H}|\Phi\rangle$ and $e^{-i \Delta\beta \hat A}|\Phi\rangle$ and fidelity of QITE with respect to the exact ground state for N$_2$ at the equilibrium bond length (left) and stretched bond length (right).} \label{fig:F}
\end{figure}

With the present equation (\ref{eq:b}), which is free from the squared norm $c$, the energy convergence is almost independent of the selected $\Delta\beta$ (blue curves). Therefore, it is permitted to use a large $\Delta\beta$ to reduce the number of times to evaluate {\bf a}; in Figs.~\ref{fig:gs}(d), (e), and (f), the result with $\Delta\beta=0.1$ provides the fastest convergence in terms of the number of $\ell$. 

These behaviors can be also explained by Fig.~\ref{fig:F}, which depicts $F({\bf a})$ of Eq.~(\ref{eq:F}) and the fidelity of QITE state with respect to the exact state using N$_2$ with $\Delta\beta = 0.1$ {\it a.u.}. The imaginary-time-propagated state $\frac{1}{\sqrt{c}} e^{-\Delta\beta\hat H}|\Phi\rangle$ is obtained by the Taylor expansion until its norm is converged. It is evident that the present derivation successfully creates a state with fidelity higher by an order of magnitude. Since $E \approx -108$ Hartree and $\Delta\beta = 0.1$, the update coefficients {\bf a} are scaled by $\frac{1}{\sqrt{c}} \approx 0.2$ in the previous algorithm. This results in the slow-down of the fidelity with respect to the exact state,  especially for the strongly correlated dissociating N$_2$.

All these results demonstrate the correctness of our derivation. 
We note that the convergence of QITE becomes slower for strongly correlated systems, even with the corrected equation. The slow-down can be attributed to the fact that the initial HF state has a rather small overlap with the final state, and can be mitigated by the use of a multi-determinant state instead of HF.

\subsection*{Truncation of operator basis in Fermionic-QITE}
While the UCCGSD {\it ansatz} includes all possible single and double excitation operators, it is anticipated that many operators are irrelevant for the unitary approximation of the imaginary time propagation. Hence, we consider to use only the fermion operators that appear in $\hat H$ to reduce the number of Pauli operators in the QITE simulation. This ``Hamiltonian-based QITE'' is in many cases nearly equivalent to the UCCGSD-based QITE. However, it not only effectively captures the symmetry in $\hat H$ by naturally avoiding the symmetry-forbidden excitations but also provides an opportunity to screen operators by removing excitations with small integrals $h_{pq}$ and $(pr|qs)$ below some threshold $\epsilon$, assuming that they do not play a significant role. We anticipate such screening may be beneficial for molecules that possess only approximate point-group symmetry, and especially for larger systems owning to the $1/r$ decay of $(pr|qs)$ in the localized orbital basis. 
 
 Here, we assess the effect of truncating operators in the Hamiltonian-based QITE. We choose a distorted formaldehyde as our test case with the following geometry: R$_{\rm CO} = 1.205$ {\AA} and $\angle {\rm HCO} = 121.9^\circ$, and one of the CH bond bonds is marginally stretched by 0.01 {\AA} from the experimental value of 1.111 {\AA}. Therefore, the system has a $C_s$ symmetry, instead of $C_{2v}$ of the equilibrium geometry. We mapped the highest eight HF orbitals (with eight electrons) onto qubits. For the UCCSD- and UCCGSD-based QITE methods, the point-group symmetry was taken into account to make a fair comparison.

Table \ref{tb:ch2o} lists the number of Pauli operators used and final energy obtained with each method. Given that the exact energy is $-113.540$ 654 Hartree, the UCCSD- and UCCGSD-based QITE methods are both accurate. Although the former has an error of approximately 2 mHartree, it should be noted that the number of Pauli operators is about one third of that of the UCCGSD-based QITE. For the Hamiltonian-based QITE without truncation of operators ($\epsilon = 0$), both the number of Pauli terms and the final energy are identical to those of the UCCGSD-based QITE (reduced by considering the point-group symmetry), as expected.

\begin{table}
\caption{Performance of UCCSD-, UCCGSD-, and Hamiltonian-based QITE with operator truncation. $\epsilon$ is the truncation threshold in  choosing the operators in the Hamiltonian-based QITE.}\label{tb:ch2o}
\tabcolsep= 3mm
\begin{tabular}{llrc}
\hline\hline
  Ansatz &$\epsilon$ (a.u.) & Terms & Energy (a.u.) \\
   \hline
   UCCSD   & ---        &  1416 & $-$113.538 846\\
   UCCGSD  & ---        &  4640 & $-$113.540 649 \\
   Hamiltonian  &   0   &  4640 & $-$113.540 649 \\
		 		& 0.001 & 3172  & $-$113.540 649 \\
		 		& 0.005 & 1980  & $-$113.540 643 \\
		 		& 0.01  & 1604  & $-$113.540 635\\
		 		& 0.02  & 1068  & $-$113.540 318\\
	     		& 0.03  &  824  & $-$113.539 848\\
		 		& 0.04  &  628  & $-$113.536 544\\
		 		& 0.05  &  432  & $-$113.534 146\\
		 		& 0.1   &  160  & $-$113.517 881 \\
      \hline\hline
\end{tabular}	
\end{table}

\begin{figure}
	\includegraphics[width=80mm]{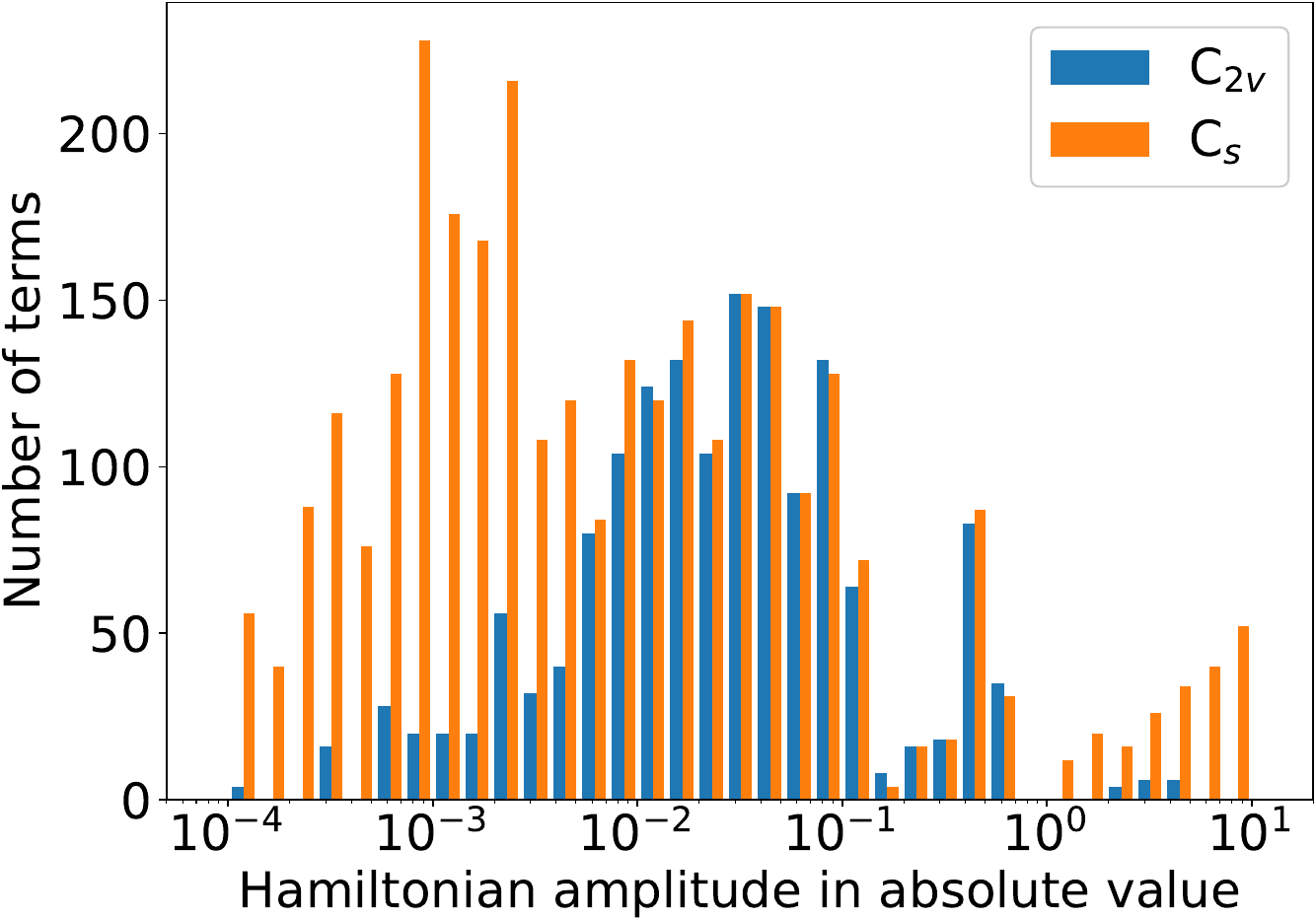}
	\caption{Number of terms in each Hamiltonian for CH$_2$O as a function of the magnitude of amplitudes.} \label{fig:terms}
\end{figure}

The advantage of the Hamiltonian-based QITE is that it contains the information about the system such as locality and approximate symmetry. In the present case, the geometry is approximately $C_{2v}$, but because of the small distortion we have introduced, the $C_s$ Hamiltonian has several small terms that are not present in the $C_{2v}$ Hamiltonian. This is illustrated in Fig.~\ref{fig:terms}, where the number of terms in each Hamiltonian is plotted as a function of the magnitude of amplitudes. The total numbers of terms in the $C_{2v}$ and $C_s$ Hamiltonians are 1545 and 3057, respectively. It is seen in Fig.~\ref{fig:terms} that most of the additional terms in the $C_s$ Hamiltonian are small (with coefficients less than 0.005). These terms are not expected to play a significant role in the propagation and hence their anti-Hermitian operators are considered unimportant in constructing $\hat A$. In fact, if we remove these operators in the Hamiltonian-based QITE with $\epsilon = 0.005$, the energy remains almost unaffected although  the number of terms is reduced by more than half (see Table~\ref{tb:ch2o}). One can further increase the threshold to $\epsilon = 0.01$ while attaining the 0.01 mHartree error.

The Pauli operators in the Hamiltonian-based and UCCGSD-based methods can be overcomplete because some of them may act trivially to an arbitrary state, or may be able to create the (nearly) same state in several different ways. 
Therefore, the good performance of the truncation scheme in the Hamiltonian-based QITE method is attributed to the fact that removing some Pauli operators simply results in the reduction of the degree of redundancy to some extent, which should not affect the accuracy in the case of a relatively small $\epsilon$. 
For a larger $\epsilon$, however, the energy becomes inaccurate with fewer terms, neglecting the essential excitation operators.
Therefore, as shown in Table~\ref{tb:ch2o}, a trade-off exists between the reduction in the number of terms and the accuracy in the resulting energy. 


\begin{figure*}
\includegraphics[width=140mm]{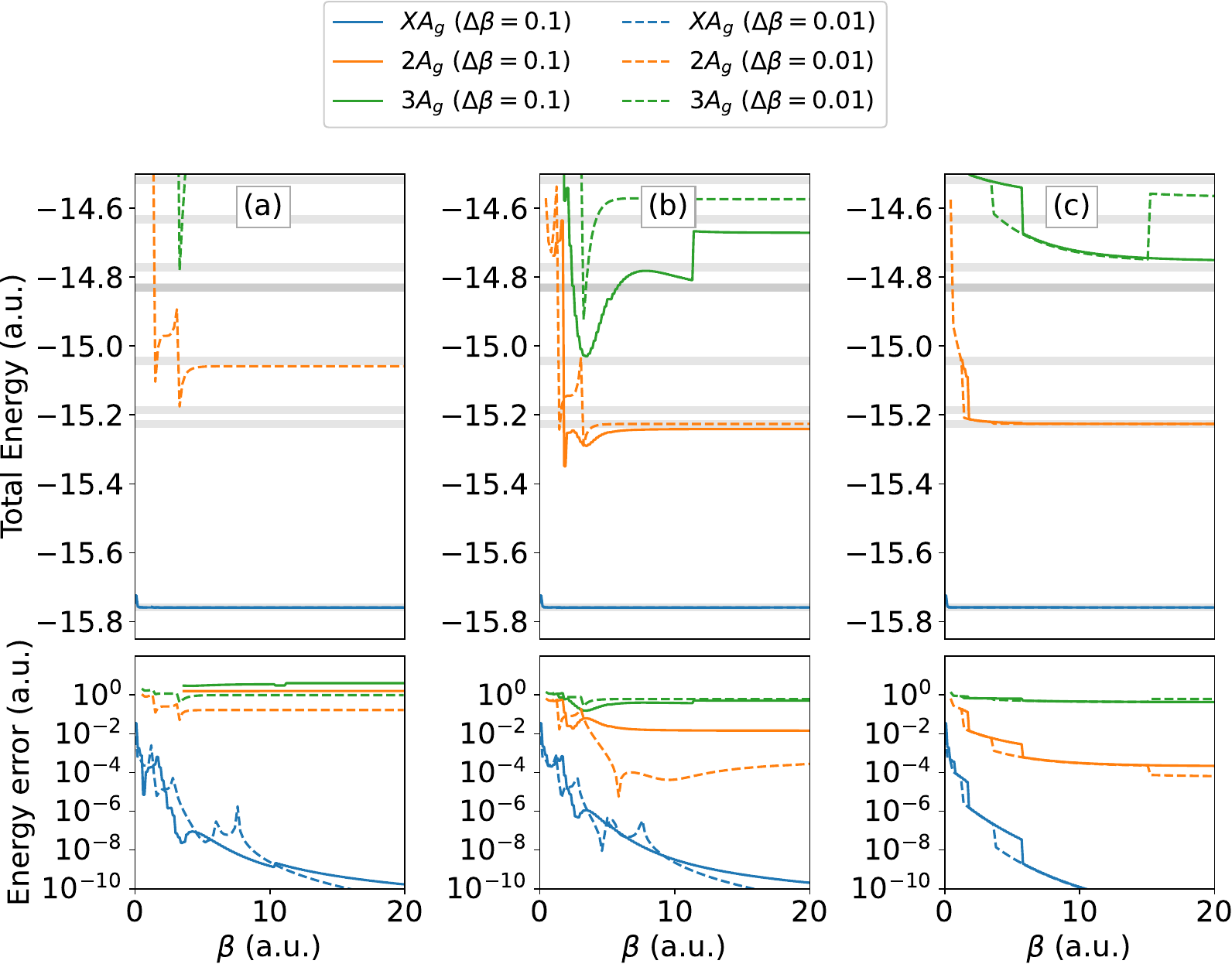}
\caption{Excited states by QLanczos using different approximations: (a) first-order approximation for the squared norm $c$, (b) shifted approximation  for the squared norm $c$, (c) exact ITE. Top and bottom panels represent the total energy profiles and errors from the exact values.}\label{fig:qlanczos}
\end{figure*}

\subsection*{Excited state calculations using QLanczos}
Let us now turn our attention to excited states. 
As pointed out by other authors\cite{Yeter-Aydeniz20}, QITE is expected to reach some excited state $|\psi_1\rangle$ by starting from $|\Phi_1^{(0)}\rangle$ that is a zeroth-order state with a large overlap, if it is orthogonal to the ground state: $\langle \Phi_1^{(0)} | \psi_g\rangle = 0$. This condition is usually satisfied by the different symmetries in the wave functions $|\psi_1\rangle$ and $|\psi_0\rangle$. Therefore, in principle, one can simulate excited states by QITE, provided that they are the lowest state of some irreducible symmetry. However, when the symmetry cannot be exploited (i.e., the target excited state has the same symmetry as the ground state), the approach always converges to the ground state by virtue of ITE. In the Supplementary Information, we discuss the role of symmetry in excited state calculations. 
 
Therefore, an interesting question to be answered is whether QLanczos could provide some excited states as higher eigenstates in general (note that its subspace only contains states with the same symmetry). 
It should be first noted that, in QLanczos, the energy error source is mainly four-fold: (i) the approximation of $c^{(\ell)}$  at each time step, (ii) the unitary approximation introduced in the QITE algorithm, (iii) the consequence of the truncation of the Krylov basis due to the linear dependence in the generalized eigenvalue problem, and (iv) noise in quantum devices (for the detail of QLanczos algorithm, see the Methods section below and Ref.\cite{Motta20}). 

Here, we mainly focus on the error arising from the approximation (i). Many previous work have approximated $c^{(\ell)}$ to first order,
 \begin{align}
 c^{(\ell)} &= \langle \Phi^{(\ell)}| e^{-2\Delta\beta \hat H} |\Phi^{(\ell)}\rangle \nonumber\\
 &= 1 - 2\Delta\beta \langle \Phi^{(\ell)}|\hat H|\Phi^{(\ell)}\rangle  + O(\Delta\beta^2)  \label{eq:c_approximation1}
 \end{align}
Sun et al. employed the second-order approximation\cite{Sun21}. However, these approximations suffer an exponential error with the energy increase (larger systems), entailing small $\Delta\beta$.

In order to evaluate $c^{(\ell)}$ more appropriately, it is desirable to consider the energy shifted propagator, namely, we write
\begin{align}
	c^{(\ell)} &= e^{-2\Delta\beta E^{(\ell)}}\langle \Phi^{(\ell)} | e^{-2\Delta\beta (\hat H - E^{(\ell)})}|\Phi^{(\ell)}\rangle \label{eq:c_approximation2.0}
\end{align}
the expansion of which converges much faster if we define $E^{(\ell)} = \langle \Phi^{(\ell)} | \hat H | \Phi^{(\ell)}\rangle$. The factor $e^{-2\Delta\beta E^{(\ell)}}$ is exactly computable if $E^{(\ell)}$ is measured, and can be set to a constant for the purpose of first-order expansion of the expectation value part of Eq.~\ref{eq:c_approximation2.0}. Thus, we find
\begin{align}
	c^{(\ell)} &= e^{-2\Delta\beta E^{(\ell)}} \Big(1 + 2 \Delta\beta^2 \langle\Phi^{(\ell)}|(\hat H - E^{(\ell)})^2  |\Phi^{(\ell)}\rangle \nonumber\\
	&+ O(\Delta\beta^3)\Big)\nonumber\\
	& \approx e^{-2\Delta\beta E^{(\ell)}}\label{eq:c_approximation2}
\end{align}
where the second-order and higher terms rapidly decay, and the first-order approximation remains reasonable, especially when we are in the vicinity of the convergence.

To give some explicit numbers, here we consider the same BeH$_2$ molecule system as above. At the initial time $\ell = 0$ (i.e., with the HF state), $c^{(0)} = \langle \Phi^{(0)} | e^{-2\Delta\beta \hat H} |\Phi^{(0)}\rangle$  is evaluated to be 23.235 {\it a.u.} if exactly calculated with $\Delta\beta = 0.1$ {\it a.u.}. Using the first-order approximation~(\ref{eq:c_approximation1}), we have $c^{(0)} \approx 4.145$ {\it a.u.}, given that the HF energy is $-15.7240$ Hartree, which is completely inadequate for use (although this large error is somewhat canceled out and mitigated in the evaluation of matrix elements, see Eq.~(\ref{eq:Smat}) in the Methods section). In contrast, the shifted approximation (\ref{eq:c_approximation2}) yields $c^{(0)} \approx 23.215$ {\it a.u}, which is more appropriate. Note that as $\beta$ becomes large, the error becomes smaller in Eq.~(\ref{eq:c_approximation2}) because the QITE state better approximates the eigenstate of $\hat H$, whereas that in Eq.~(\ref{eq:c_approximation1}) remains large. Now, how do these approximations of $c^{(\ell)}$ affect the results of excited state calculations in QLanczos?

Fig.~\ref{fig:qlanczos} depicts the total energies and errors of the ground and two excited states (of the $A_g$ point-group symmetry) obtained by different approximations in QLanczos: Fig.~\ref{fig:qlanczos}(a) and (b) use the first-order approximation and the energy-shifted approximation, respectively. On the top panels, the energy changes are plotted along with the exact energies (grey lines). We have used the UCCGSD-based QITE.

It is evident that the accuracy of $c^{(\ell)}$ critically affects the accuracy of excited states in QLanczos although that of the ground state $XA_g$ is almost independent of it, owning to the variational nature of the lowest ground state.
Using the straightforward first-order approximation of $c$ with $\Delta\beta = 0.1$ {\it a.u.}, the second and third eigenvalues of QLanczos converge to $-13.6085$  and $-11.1386$ Hartree at $\beta =20$ {\it a.u.}, which are both way higher than the true values, $-15.2263$ and $-15.1858$ Hartree (see Fig.~\ref{fig:qlanczos}(a)). These values are slightly improved to $-15.0586$ and $-14.2095$ Hartree when the time step is made finer to $\Delta\beta = 0.01$ {\it a.u.}, which permits a better approximation of $c$. If, instead, we use the energy-shifted $c$ of Eq.~(\ref{eq:c_approximation2}), we obtain $-15.2408$ and $-14.6710$ Hartree for $\Delta \beta = 0.1$ {\it a.u.} and $-15.2261$ and $-14.5735$ Hartree for $\Delta\beta = 0.01$ {\it a.u.}. The error in the first excited state is substantially reduced, while the second excited state is still not captured correctly. 

The large error in the second excited state is not attributed to the approximation in $c$, but rather to the truncation of the used Krylov space. To see how important these effects are, we have also performed QLanczos with the {\it exact} imaginary time evolution in Fig.~\ref{fig:qlanczos}(c). As expected, the convergence profiles remain mostly unchanged regardless of the different $\Delta\beta$ because $c$ is exactly evaluated. The error in the second excited state comes from the error source (iii), since everything else is treated exactly; the numerical linear dependence in the basis has to be removed  and the discarded Krylov space turns out to contain important components of excited states. Note that this also causes the small energy discrepancy in the first excited state (about $10^{-4}$ Hartree at $\beta = 20$ {\it a.u.}). The errors of exact QLanczos are quite similar to those with the energy-shifted $c$ and $\Delta\beta = 0.01$ {\it a.u.} in Fig.~\ref{fig:qlanczos}(b), indicating Eq.~(\ref{eq:c_approximation2}) is fruitful. 

In conclusion, these results confirm that the components of the excited states in the qubits tend to diminish to zero, and QLanczos's ability to estimating excited states becomes less effective for higher energy states.


\subsection*{Folded-spectrum QITE and QLanczos}
To obtain high-lying states with QITE, it is required to retain their components during the imaginary time evolution. One straightforward way to achieve this is to employ the following propagator,
\begin{align}
	\hat P = e^{-\beta^2 (\hat H - \omega)^2}\label{eq:FS}
\end{align}
where $\omega$ is a target energy\cite{Booth12}. This approach originates from the folded spectrum (FS) method that was also used in the context of VQE\cite{McClean16,Santagati18, Zhang21A}. Eq.~(\ref{eq:FS}) in principle projects out the exact excited state that has an energy close to $\omega$ in the limit $\beta^2 \rightarrow \infty$. Implementing FSQITE requires little modification to the existing algorithm; namely, one can simply replace $\hat H$ in QITE with $(\hat H - \omega)^2$.

\begin{figure}
\includegraphics[width=80mm]{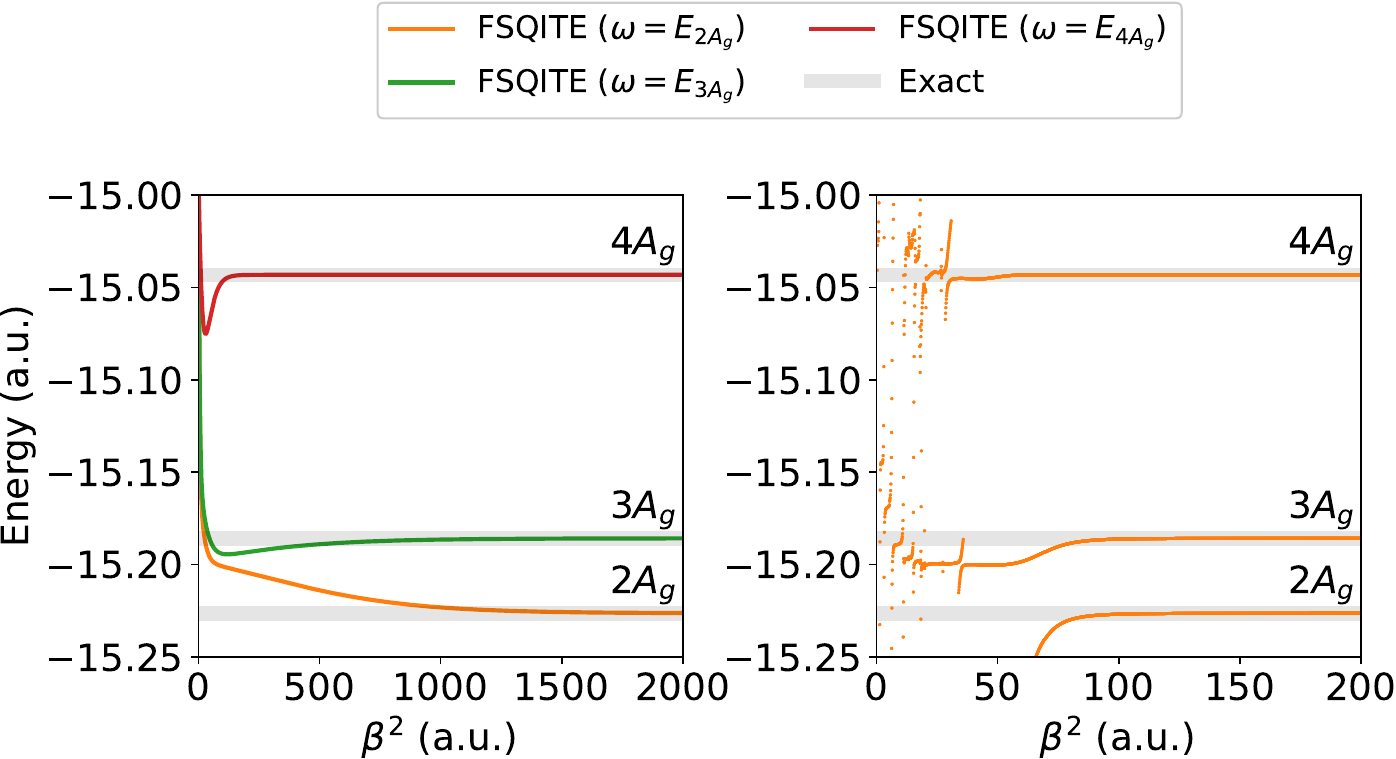}
\caption{FSQITE results for BeH$_2$ excited states. (a) FSQITE energies with different target energies $\omega$. (b) FS-QLanczos energies with $\omega=E_{2A_g}$.}\label{fig:BeH2_fsqite}
\end{figure}

In Fig.~\ref{fig:BeH2_fsqite}(a), plotted are the energy changes of three FSQITE simulations conducted for the BeH$_2$ system, targeting the $2A_g$, $3A_g$, and $4A_g$ excited states with $\omega = E_{2A_g}$, $E_{3A_g}$, and $E_{4A_g}$, respectively (i.e., the exact excited state energies). The initial configuration used for these simulations is the one with two electrons promoted from the highest occupied orbital to the lowest vacant orbital (see Supplementary Fig.~S1(c)), and $\Delta \beta^2 =0.05$ {\it a.u.} is employed for the time step. It is striking that, although FSQITE does find the desired excited state, in some cases it experiences a considerably slow evolution despite the use of the ideal target energies; it takes $\beta^2 = 1370$, $870$, and $130$ {\it a.u.} to reach the accuracy within the 1 mHartree error from the converged energy, for the $2A_g$, $3A_g$, and $4A_g$ states. 

Therefore, it is highly desirable to accelerate the convergence. Since the algorithmic difference between QITE and FSQITE lies only in the propagator form, QLanczos can be applied directly even in FSQITE. In Fig.~\ref{fig:BeH2_fsqite}(b), we present the QLanczos results using $\omega=E_{2A_g}$ (and the energy-shifted approximation for $c$). Such FS-QLanczos provides a drastic speed-up in capturing excited states from FSQITE. Although FS-QLanczos still requires $\beta^2\approx 100$ {\it a.u.}, FS-QLanczos is able to capture higher excited states, $3A_g$ and $4A_g$. This behavior can be attributed to the use of the folded-spectrum propagator Eq.~(\ref{eq:FS}), which projects the non-dominant states at a slower rate than the standard propagator, thus holding the information about other excited states nearby the target one. 

Having seen the good performance of FSQITE especially when combined with QLanczos, we should also point out its drawbacks. First, the obvious one is that $\omega$ must be specified in advance. Second, dealing with $\hat H^2$ is rather demanding. Third, it will face a difficulty in nearly-degenerate excited states. Finally, the Trotter step $\Delta\beta^2$ needs to be one or two orders of magnitude smaller than $\Delta\beta$; in other words, the simulation may take considerably longer than the ground state QITE. However, when these difficulties are overcome, FSQITE and FS-QLanczos can be promising strategies for excited states.

\section*{Discussion}\label{sec:conclusion}
In this work, we proposed several extensions to quantum imaginary time evolution with respect to their applications that include excited states. We especially focused on chemical Hamiltonians, where the Hamiltonian-based QITE was found to be successful, taking an advantage of the Hamiltonian under consideration. It contains the anti-Hermitian operators created from the local Hamiltonian terms. It also takes into account approximate point-group symmetry with the operator truncation, while maintaining the accuracy. 

Furthermore, we showed that the newly derived equation for QITE outperforms the original algorithm for the ground states of larger systems, by eliminating the dependence on the normalization constant. The original derivation entails a long imaginary time evolution for systems with a large energy, whereas our algorithm has proved to be robust with the use of larger time steps and thus can save quantum resources. 

This work also discussed how to obtain excited states. Our simulations revealed that QLanczos may be able to extract the lowest excited state if the squared norm is correctly estimated (especially with small $\Delta\beta$), but higher energy states are not found as they are discarded in the orthonormalization process of QLanczos. FSQITE offers a way to approaching arbitrary excited states that QLanczos by itself cannot finds. Its slow convergence is largely mitigated by combining it with QLanczos. 

We hope that our findings provide new insights to the quantum computing community, and that they benefit the further developments of advanced quantum algorithms.

\section*{Methods}
\subsection*{Equations for QLanczos }\label{app:QLanczos}
In QLanczos, the eigenstates of the Hamiltonian are expanded by the time-evolved states, $\left\{|\Phi^{(\ell)}\rangle\right\}$\cite{Motta20}. The Hamiltonian matrix elements are given by
\begin{align}
{\mathscr H}_{\ell, \ell'} = \langle \Phi^{(\ell)}| \hat H|  \Phi^{(\ell')} \rangle =  {\mathscr S}_{\ell, \ell'} E^{(\frac{\ell+\ell'}{2})} 
\end{align}
where the overlap matrix elements ${\mathscr S}_{\ell, \ell'}$ are built by the following relation:
\begin{align}
	{\mathscr S}_{\ell, \ell'} = \langle \Phi^{(\ell)}|  \Phi^{(\ell')} \rangle = \frac{n^{(\ell)}n^{(\ell')}}{(n^{(\frac{\ell+\ell'}{2})})^2}\label{eq:Smat}
\end{align}
with $n^{(\ell)}$ being the normalization constant at the $\ell$th time. To evaluate this, we write
\begin{align}
	\frac{1}{(n^{(\ell+1)})^2} &= \frac{\langle \Phi^{(\ell)} |e^{-2\Delta\beta \hat H}|\Phi^{(\ell)}\rangle}{(n^{(\ell)})^2} = \frac{c^{(\ell)}}{(n^{(\ell)})^2} \nonumber\\
	&= \prod_{k=0}^{\ell} c^{(k)}\label{eq:n}
\end{align}
where $c^{(\ell)}$ is the squared norm and, in the last equality, we have recursively applied the relation with $n^{(0)} = 1$. After a simple algebra, this gives
\begin{align}
	{\mathscr S}_{\ell, \ell'} = \prod_{k=1}^{\ell -\ell'}\sqrt{\frac{c^{(\ell'+k-1)}}{c^{(\ell-k)}}}
\end{align}
Then one solves a generalized eigenvalue problem
\begin{align}
{\bm {\mathscr H}} {\bf x} = {\bm {\mathscr S}}{\bf x} {\bm {\mathscr E}}\label{eq:gev}
\end{align} 
  	
Therefore, the eigenvalues ${\mathscr E}_i$ can vary depending on the approximation for $c^{(\ell)}$, and it is suggested to employ Eq.~(\ref{eq:c_approximation2}). For very large systems (with energy in the order of thousands of Hartree), Eq.~(\ref{eq:c_approximation2}) might also become unstable because of the exponential increase with $-E^{(\ell)}$. Therefore, we can further shift $E^{(\ell)}$ by some fixed reference energy $E_0$, e.g., the HF energy throughout the imaginary time evolution. Namely, one can set $\Delta E^{(\ell)} = E^{(\ell)} - E_0$ and define, instead of $c^{(\ell)}$,
\begin{align}
	\tilde c^{(\ell)} & =  e^{2\Delta\beta  E_0} c^{(\ell)}  \approx e^{-2\Delta\beta \Delta E^{(\ell)}} \label{eq:c_approximation3}
\end{align}
which is drastically smaller because $\Delta E^{(\ell)}$ is the correlation energy, which is several orders of magnitudes smaller than the total energy, $|\Delta E^{(\ell)}| \ll |E_0| < |E^{(\ell)}|$. Since we use Eq.~(\ref{eq:Smat}) to evaluate the matrix elements combined with Eq.~(\ref{eq:n}), the factor $e^{-2\Delta\beta  E_0}$ in the numerator and denominator cancel out exactly. Note that the use of Eq.~(\ref{eq:c_approximation2}) and Eq.~(\ref{eq:c_approximation3}) yield exactly the same matrix elements in arithmetic, and the advantage of the latter is just the numerical stability it offers. 

\subsection*{QLanczos stabilization}\label{app:QLanczos_stability}
To treat the linear dependence in solving Eq.~(\ref{eq:gev}), Ref.[\onlinecite{Motta20}] proposed to use only Krylov vectors that satisfy ${\mathscr S}_{\ell \ell'} < s$ with some threshold $s$. However, we found that this procedure often fails to find excited states. Therefore, we use the whole Krylov space; however, the linear dependence introduces numerical instabilities and makes it difficult to interpret the results. To circumvent this problem, in our algorithm, the physical solutions of QLanczos are identified by looking at the eigenvectors in the orthogonal space. This procedure is described as follows.

First, we diagonalize the overlap matrix ${\bm{\mathscr S}}$ to obtain the eigenvalues $\eta_i$. These eigenvalues indicate the importance of the basis, i.e., the larger the value is, the more relevant it is to the physical space, and {\it vice versa}. Once the L\"owdin orthonormalization vectors are obtained, each of them is classified as either physically relevant one or irrelevant one (redundant), based on the magnitude of the corresponding eigenvalue. Here, we consider the vector is physically relevant if $\eta_i > 0.01$. The vectors with $\eta_i < 10^{-8}$ are removed. Note that, since $\eta_i$ decay exponentially, only one or two vectors are relevant in several cases, making it challenging to determine higher energy states.

Then, we diagonalize the QLanczos Hamiltonian appropriately orthogonalized within this subspace, 
\begin{align}
	{\bm{\mathscr H}}_{\rm ortho} {\bf v}^{(k)} = {\mathscr E}_k {\bf v}^{(k)} 
\end{align} We identify the physically meaningful energies by inspecting the eigenvectors ${\bf v}^{(k)}$; if the physically relevant component $v^{(k)}_i$ is larger than 
some threshold $|v^{(k)}_i|^2>\varepsilon$, which we took as 0.1 in this work, then the $k$th eigenvalue is considered to represent either the ground or excited state.

\section*{Data Availability}
The data that support the findings of this study are available from the corresponding author upon reasonable request.

\section*{Code Availability}
The code that is used to produce the data presented in this study is available from the authors upon reasonable request.

\section*{Acknowledgements}
This work was supported by JST, PRESTO (Grant Number JPMJPR2016), Japan and  by JSPS KAKENHI (Grant Number JP20K15231). We are grateful for the computational resources provided by ECCSE, Kobe University.

\section*{Author contributions}
T.T. conceived the idea and wrote the paper. Y.R. and T.T. implemented the algorithms and performed numerical simulations. T.T., Y.R., and S.L.T. all participated in discussions that developed the theory and shaped the project.


%
\end{document}